\documentclass[12pt]{article}
 \usepackage{graphicx}
 \usepackage[cp1251]{inputenc} 
\begin{document}
 \noindent {\footnotesize\it Astronomy Letters, 2016, Vol. 42, No. 4, pp. 228--239.}
 \newcommand{\dif}{\textrm{d}}

 \noindent
 \begin{tabular}{llllllllllllllllllllllllllllllllllllllllllllll}
 & & & & & & & & & & & & & & & & & & & & & & & & & & & & & & & & & & & & & \\\hline\hline
 \end{tabular}

  \vskip 0.5cm
  \centerline{\bf Hercules and Wolf 630 Stellar Streams and Galactic Bar Kinematics}
  \bigskip
  \centerline{V.V. Bobylev and A.T. Bajkova}
  \bigskip
  \centerline{\small\it Pulkovo Astronomical Observatory, St. Petersburg,  Russia}
  \bigskip
  \bigskip
{\bf Abstract}—We have identified the four most significant
features in the $UV$ velocity distribution of solar neighborhood
stars: H1, H2 in the Hercules stream and W1, W2 in the Wolf 630
stream. We have formulated the problem of determining several
characteristics of the central Galactic bar independently from
each of the identified features by assuming that the Hercules and
Wolf 630 streams are of a bar-induced dynamical nature. The
problem has been solved by constructing $2:1$ resonant orbits in
the rotating bar frame for each star in these streams. Analysis of
the resonant orbits found has shown that the bar pattern speed is
45--55 km s$^{-1}$ kpc$^{-1}$, while the bar angle lies within the
range $40^\circ-60^\circ$. The results obtained are consistent
with the view that the Hercules and Wolf 630 streams could be
formed by a long-term influence of the Galactic bar leading to a
characteristic bimodal splitting of the $UV$ velocity plane.


\section*{INTRODUCTION}
Analysis of the velocity field of solar-neighborhood stars using
Hipparcos (1997) data (Chereul et al. 1998; Dehnen 1998; Asiain et
al. 1999; Skuljan et al. 1999) and based on the most recent data
(Famaey et al. 2005; Bobylev and Bajkova 2007; Antoja et al. 2008;
Bobylev et al. 2010) has revealed a well-developed fine structure.
Various nonaxisymmetric Galactic potential models, in particular,
the spiral pattern and the Galactic bar, are considered to explain
several structures to which quite old stars belong.

As simulations showed, the existence of the Hercules stream can be
explained by the fact that its stars have resonant orbits induced
by the Galactic bar (Dehnen 1999, 2000; Fux 2001; Chakrabarty
2007). A detailed analysis of the kinematics of nearby F and G
dwarfs using high-resolution spectra (Bensby et al. 2007) showed
the stars in the stream to have a wide range of stellar ages,
metallicities, and elemental abundances, and this led to the
conclusion that the dynamical effect of the Galactic bar is the
most acceptable explanation for the existence of the Hercules
stream.

On the whole, the Galactic evolution simulations performed with
various Galactic potential models and central bar models confirm
the hypothesis about a dynamical origin for some of the streams
observed in the solar neighborhood, in particular, the Hercules
stream (Gardner and Flinn 2010; Bovy 2010). The point is that the
central bar in our Galaxy can be represented not as a singular
one. At least, there can be two of them: one is long and the other
is short; they are oriented at different angles with respect to
the direction toward the Sun and can rotate with different pattern
speeds. All of this makes a detailed reproduction of the observed
stellar velocity distribution in the solar neighborhood difficult.

On the other hand, Antoja et al. (2009) showed that some of the
dynamical stellar streams in the solar neighborhood, such as the
Hercules stream, could be due to the existence of dynamical
resonances with the Galactic spiral structure and not exclusively
to the influence of the Galactic bar. As several authors showed,
the influence of spiral structure leads to a splitting of the $UV$
velocity plane and the emergence of clumpiness (De Simone et al.
2004; Quillen and Minchev 2005), but it cannot completely explain
the global asymmetry (the $U$--anomaly problem, the distribution
bimodality) in the observed $UV$ velocity distribution.

In addition to the Hercules stream, there are also other features
on the $UV$ velocity plane, in particular, the Wolf 630 stream
(Bobylev et al. 2010), with some of the authors designating this
place as the $\alpha$~Ceti stream (Francis and Anderson 2009). The
interest is that there are orbits both elongated along the bar
major axis and oriented perpendicularly to it in the bar reference
frame. At present, there is reason to believe that the Sun is near
the point of intersection of such orbits (Fig.~1 in Dehnen
(2000)). In this case, the bimodality of the $UV$ velocity
distribution can be explained by the fact that there are
representatives of these two orbit families in the solar
neighborhood. According to Dehnen (2000), the detailed form of the
velocity distribution depends strongly on the position of the
Outer Lindblad Resonance (OLR). In particular, there will be no
bimodality if the OLR is farther from the Galactic center than the
solar cycle ($R_{OLR}/R_0>1.05,$ Fig.~4 in Dehnen (2000)).

According to the coordinates on the $UV$ velocity plane, the
Hercules stream is a representative of the family of orbits
oriented perpendicularly to the bar major axis, while the
representatives of the family of orbits elongated along the bar
major axis are located in the region on the $UV$ velocity plane
where the Wolf 630 stream is observed.

Note that both the relic of an open cluster and the stars that
appeared here due to resonances or other dynamical effects can be
simultaneously present in the clump on the $UV$ velocity plane.
Such a situation is observed in the region of the Hyades stream,
where there are stars of the well-known open cluster, but stars
with resonant orbits constitute the bulk of the stream (Pompeia et
al. 2011). When studying the Wolf 630 stream based on the list of
Eggen (1969), Bubar and King (2010) found 19 stars with an
exceptional homogeneity among 34 candidates: [Fe/H]=$-0.01\pm0.02$
at a mean age of $2.7\pm0.5$~Gyr. This argues for the hypothesis
about the relic of an open cluster or a dwarf galaxy. On the other
hand, using $\sim$200 stars of the Wolf 630 stream, Bobylev et al.
(2010) showed its inhomogeneity: [Fe/H]$\approx-0.1\pm0.2$ at a
mean age of $\sim4\pm2.5$~Gyr.

The goal of this paper is to estimate such characteristics of the
bar as its pattern speed $\Omega_{bar}$ and orientation
$\theta_{bar}$ under the assumption that the Hercules and Wolf 630
streams could be formed by a single mechanism associated with the
influence of the Galactic bar. We solve the problem by
constructing $2:1$ resonant orbits of individual stars forming
these streams in the rotating bar frame and by analyzing the
Galactic rotation curve. The characteristics of each of the
streams are determined independently, i.e., we do not assume them
to be of common origin in advance.

 \begin{figure} {\begin{center}
 \includegraphics[width=95mm]{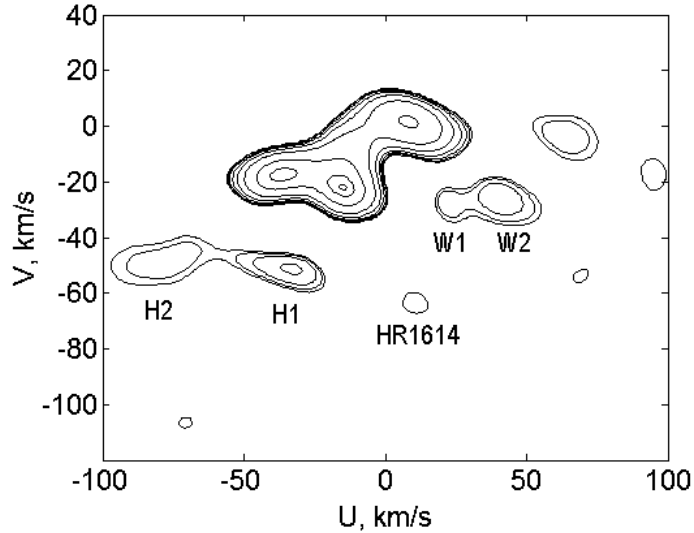}
 \caption{Wavelet map of $UV$ velocities constructed using $\sim$17000 single stars
 from Hipparcos (1997) with relative parallax errors of no more than 10\%.}
 \label{f1} \end{center} } \end{figure}

\section*{THE $UV$ VELOCITIES OF THE STREAMS}
The wavelet map of the $UV$ velocity distribution for single stars
with reliable distance estimates from Bobylev et al. (2010) is
presented in Fig.~1. The map was constructed using $\sim$17000
stars from the Hipparcos catalogue (1997). We took their proper
motions and parallaxes from a revised version of this catalogue
(van Leeuwen 2007) and their radial velocities from the OSACA
catalogue of radial velocities (Bobylev et al. 2006) and the
Pulkovo compilation of radial velocities (PCRV) (Gontcharov 2006).
For all these stars, the relative error in the parallax does not
exceed 10\%.

The contour lines in Fig. 1 are given on a logarithmic scale: 1,
2, 4, 8, ..., 90\%, and 99\%. The velocities are given relative to
the Sun. This figure indicates the W1 and W2 features for the Wolf
630 stream and the H1 and H2 features for the Hercules stream with
the following coordinates of their centers: W1 $(U,V)=(21,-26)$ km
s$^{-1}$, W2 $(U,V)=(40,-24)$ km s$^{-1}$, H1 $(U,V)=(-33,-51)$ km
s$^{-1}$, H2 $(U,V)=(-71,-48)$ km s$^{-1}$. According to this map,
we selected the specific stars belonging to these features based
on the probabilistic method. The number of probable candidates in
each stream was the following: 250, 271, 401, and 218 stars in W1,
W2, H1, and H2, respectively.

When constructing the Galactic orbits to determine the stellar
velocities relative to the local standard of rest (LSR), we use
the peculiar solar velocity components relative to the LSR with
their values from Sch\"onrich et al. (2010), $(U_\odot,
V_\odot,W_\odot)=(11.1, 12.2, 7.3)\pm(0.7, 0.5, 0.4)$ km s$^{-1}$.
When these velocities were determined, the stellar metallicity
gradient in the Galactic disk and the effect of radial mixing of
stars in the disk were taken into account and modeled. Therefore,
the values of these components are currently the most plausible
ones and are widely used by various authors.

\section*{METHODS}
\subsection*{Orbit Construction}
We construct the orbits by solving the following system of
equations of motion based on a realistic Galactic gravitational
potential (Fern\'andez et al. 2008):
 \begin{equation}
 \begin{array}{lll}
 \ddot{\xi}=-\frac{\displaystyle\partial\Phi}{\displaystyle\partial\xi}-
   \Omega^2_{\odot}(R_{\odot}-\xi)-2\Omega_{\odot}\dot{\eta},\\
 \ddot{\eta}=-\frac{\displaystyle\partial\Phi}{\displaystyle\partial\eta}+
    \Omega^2_{\odot}\eta+2\Omega_{\odot}\dot{\xi},\\
 \ddot{\zeta}=-\frac{\displaystyle\partial\Phi}{\displaystyle\partial\zeta},
 \label{1}
 \end{array}
\end{equation}
where $\Phi$ is the Galactic gravitational potential; the
coordinate system $(\xi,\eta,\zeta)$ centered on the Sun rotates
around the Galactic center with a constant angular velocity
$\Omega_{\odot}$, with the $\xi$ axis being directed to the
Galactic center, the $\eta$ axis pointing in the direction of
Galactic rotation, and the $\zeta$ axis being directed toward the
north Galactic pole; $R_{\odot}$ is the Galactocentric distance of
the Sun. When the Galactic potential $\Phi$ is known, the system
of equations~(1) can be solved numerically. We used a fourth-order
Runge–Kutta integrator.

In the model of Allen and Santill\'an (1991), the Galactocentric
distance of the Sun is taken to be $R_\odot=8.5$~kpc, while the
circular velocity of the Sun around the Galactic center is
$V_\odot=\Omega_{\odot}R_\odot=220$~km s$^{-1}$. The Galactic
potential considered here consists of an axisymmetric component
and a bar potential:
 \begin{equation}
 \Phi = \Phi_o +\Phi_b. \label{e0}
 \end{equation}
In turn, the axisymmetric component can be represented as the sum
of three components --- central (bulge), disk, and halo ones:
 \begin{equation}
 \Phi_o = \Phi_C + \Phi_D + \Phi_H.
 \label{potential}
\end{equation}
{\it The central component} of the Galactic potential in
cylindrical coordinates ($r,z,\theta$) is represented as
 \begin{equation}
 \Phi_C=-\frac{M_C}{(r^2+z^2+b^2_C)^{1/2}},
 \label{2}
 \end{equation}
where $M_C$ is the mass, $b_C$ is the scale length, and
$r^2=x^2+y^2.$ {\it The disk component} is
 \begin{equation}
 \Phi_D=-\frac{M_D}{\{r^2+[a_D+(z^2+b_D^2)^{1/2}]^{2}\}^{1/2}},
 \label{3}
\end{equation}
where $M_D$ is the mass, $a_D$ and $b_D$ are the scale lengths.
{\it The halo component} is
 \begin{equation}
 \Phi_H=-\frac{M(R)}{R}-\int_R^{100}{{\frac{1}{R^{'}}}{\frac{dM(R^{'})}{dR^{'}}}}dR^{'},
 \label{4}
 \end{equation}
where
$$
M(R)=\frac{M_H(R/a_H)^{2.02}}{1+(R/a_H)^{1.02}},
$$
Here, $M_H$ is the mass, $a_H$ is the scale length, and
$R^2=x^2+y^2+z^2.$ The triaxial ellipsoid model (Palou$\breve{s}$
et al. 1993) was chosen as the potential due to the central bar:
\begin{equation}
  \Phi_b = \frac{M_b}{(q_b^2+x^2+[ya_b/b_b]^2+[za_b/c_b]^2)^{1/2}},
\label{bar}
\end{equation}
where $x=R\cos\vartheta, y=R\sin\vartheta$, $a_b, b_b, c_b$ are
the three bar semiaxes, $q_b$ is the bar length;
$\vartheta=\theta-\Omega_{bar}t-\theta_{bar}$,
$\tan\theta=\eta/(R_{\odot}-\xi)$ ($\Omega_{bar}$ is the bar
pattern speed, $t$ is the integration time, $\theta_{bar}$ is the
bar angle relative to the Galactic $X$ and $Y$ axes).

If $R$ is measured in kpc, $M_C,M_D,M_H, M_b$ are in units of the
Galactic mass $(M_G)$ equal to $2.32\times107M_\odot,$ then the
gravitational constant $G=1$ and 100 km$^2$ s$^{-2}$ is the unit
of measurement of the potential $\Phi$ and its individual
components (4)--(7). The parameters of the model potential from
Allen and Santill\'an (1991) and the bar potential adopted here
are given in Table~1.

We pass from the moving coordinate system $(\xi,\eta,\zeta)$ to
the Galactic one $(X,Y,Z)$ using the formulas
 \begin{equation}
 \begin{array}{lll}
  X=(R_{\odot}-\xi)\cos(t\Omega_{\odot})-\eta\sin(t\Omega_{\odot}),\\
  Y=(R_{\odot}-\xi)\sin(t\Omega_{\odot})+\eta\cos(t\Omega_{\odot}),\\
  Z=\zeta.
 \label{5-6}
 \end{array}
\end{equation}
In the case of integration by the Runge–Kutta method, it should be
kept in mind that $t=0.001/1.023$ corresponds to one million years
at the distance and mass measurements adopted above. The orbit
integration time was chosen to be 4~Gyr, given that the
characteristic lifetime of the bar (from the beginning of its
formation to its destruction) is typically 2--4 Gyr. Note that
allowance for the interaction with other galaxies would be
required on long integration time scales.

 {\begin{table}[t]                                                
 \caption[]
 {\small\baselineskip=1.0ex Parameters of the Galactic potential model}
 \label{t:1}
 \begin{center}\begin{tabular}{|c|c|}\hline
 $M_C$ &  606.0 M$_G$ \\
 $M_D$ & 3690.0 M$_G$ \\
 $M_H$ & 4615.0 M$_G$ \\
 $M_b$ & 43.1   M$_G$ \\
 $b_C$ & 0.3873 kpc   \\
 $a_D$ & 5.3178 kpc   \\
 $b_D$ & 0.2500 kpc   \\
 $a_H$ &   12.0 kpc   \\
 $q_b$ &    5.0 kpc   \\
 $a_b/b_b$& $1/0.42$  \\
 $a_b/c_b$& $1/0.33$  \\\hline
 \end{tabular}\end{center}\end{table}}

\subsection*{Searching for Periodic Orbits}
This section is devoted to searching for resonant periodic orbits
in an axisymmetric potential in a rotating frame lying in the
Galactic plane centered at the Galactic center.

Note that the objects execute a three-dimensional motion in
accordance with Eqs. (3)--(6) for the potential. However, since
the bar rotates around the Galactic center in the $(X,Y)$ plane
and since resonant orbits are searched for precisely in the bar
rotation frame, we search for periodic orbits in the $(Xp,Yp)$
plane (Fux 2001). We pass from the moving coordinate system
$(\xi,\eta)$ to the coordinate system $(X_p,Y_p)$ rotating with
the pattern speed $\Omega_p$ using formulas similar to Eqs. (8):
 \begin{equation}
 \begin{array}{lll}
  X_p=(R_{\odot}-\xi)\cos(t(\Omega_{\odot}-\Omega_p))-\eta
   \sin(t(\Omega_{\odot}-\Omega_p)),\\
  Y_p=(R_{\odot}-\xi)\sin(t(\Omega_{\odot}-\Omega_p))+\eta
   \cos(t(\Omega_{\odot}-\Omega_p)).
 \label{7-8}
 \end{array}
\end{equation}
In this paper, as the rotating frame we consider the bar whose
pattern speed and orientation we attempt to estimate by
constructing the resonant periodic orbits of stars from the
Hercules and Wolf 630 streams in the bar frame. This requires
finding the periodic orbits that satisfy the $(2:1)$ resonance
(Fux 2001; Orlov and Sotnikova 2007). The condition for a
resonance in the stellar disk plane is the commensurability of two
frequencies: the angular velocity $\Omega$ and the epicyclic
frequency $\kappa$:
 \begin{equation}
 l\kappa=m(\Omega-\Omega_p),
  \label{9}
 \end{equation}
with integers $m\ge 0$ and $l$. In the rotating frame, the $m/l$
resonant orbit is closed after $m$ radial oscillations and $|l|$
orbital periods. Note that condition (10) coincides with the
condition for Lindblad resonances only at $l=\pm1.$ The outside
corotation, $l<0,$ is negative, and we can talk about the outer
$m/|l|$ resonances. In this paper, we discuss only the outer
resonances. Figure~4 from Fux (2001) presents such orbits at
$|l|=1.$ If $m=2$ and $l=-1,$ corresponding to $(2:1)$ resonant
orbits, then it follows from Eq.~(10) that
 \begin{equation}
 \Omega_p=\Omega+\kappa/2.
 \label{99}
 \end{equation}
The ideology of the proposed method is that we seek for the bar
pattern speed $\Omega_p$ (it is different for different stars) at
which the orbit becomes resonant and, hence, periodic (in the case
of an axisymmetric potential) and quasi-periodic (in the case of a
nonaxisymmetric potential, for example, when the bar is included,
see below). We propose the following numerical algorithm of
seeking for such a pattern speed $\Omega_p$ lying within the range
$[\Omega_1,\Omega_2]$ predetermined empirically:

(1) For a given frequency range, we specify a grid of discrete
frequencies $\Omega_1+m\Delta\Omega, m=0,...,M$, where
$\Delta\Omega=(\Omega_2-\Omega_1)/M$, $M$ is an integer large
enough to obtain the required dense grid of frequencies.
Experience shows that an acceptable accuracy of the algorithm is
achieved at $\Delta\Omega=0.01$ km s$^{-1}$ kpc$^{-1}$.

(2) We integrate the stellar orbit in the coordinate system
$(\xi,\eta)$ for a sufficiently long time (several billion years
or tens of revolutions around the Galactic center).

(3) We pass to the coordinate system $(X_p,Y_p)$ by successively
substituting all of the frequencies from the grid specified above
into Eq. (9) as $\Omega_p$.

(4) We superimpose a discrete grid on the $(X_p,Y_p)$ plane with
the same step $\Delta x=\Delta y=\Delta$ in the $X$ and $Y$
directions that is chosen empirically to be small enough to
provide the required orbit reproduction accuracy. The product
$\Delta \times N$, where $N$ is the maximum number of cells in
both $X$ and $Y$ directions, gives the linear size of the
$(X_p,Y_p)$ map. In our case, this size at $N=512$ and the
discretization step $\Delta=46.875$~pc is 24~kpc.

(5) On the discrete $N\times N$ $(X_p,Y_p)$ plane, we assign ``1''
to the coordinates of the cell through which the orbit passes and
``0'' to all the remaining discrete coordinates. If the orbit
crosses a given cell several times, then we anyway assign ``1'' to
the corresponding point on the plane only once.

(6) For all maps, we count the number of ones $K.$ The map with
the smallest number of ones corresponds to the orbit with a
resonance frequency $\Omega_1+i\Delta\Omega$, where $i$ is the map
number.

(7) For the orbit found, we additionally check whether it belongs
to a certain type of resonant orbits (corotation, $1:1, 2:1,$
etc.).

Stable periodic orbits give a filling in the form of a trajectory
whose shape does not change with increasing number of revolutions
around the Galactic center. If several stable periodic orbits fall
within a given frequency range, then the algorithm chooses the
orbit with the smallest filling to which the orbit with the
smallest multiplicity of resonance frequencies corresponds. If the
orbit is not stable, then it progressively densely fills a certain
space on the $(X_p,Y_p)$ coordinate plane with increasing
integration time or number of revolutions around the Galactic
center, i.e., the number $K$ increases noticeably. Thus, the
change in $K$ with integration time serves as a measure of orbit
stability. Obviously, the number $K$ depends on the map cell size,
so that the numbers for a given object at different frequencies
from the frequency range $[\Omega_1,\Omega_2]$ must be compared
for discrete maps with the same discretization step.

Once the periodic orbits in the axisymmetric potential (3) had
been found, we checked how much the resonance frequencies changed
when the bar was included. For this purpose, in the vicinity of
the resonance frequency found, we refined its value using the
algorithm described above by taking into account the fact that the
bar pattern speed and angle are the parameters that enter into the
expression for the bar potential in accordance with Eq. (7). As
the bar angle we used the estimate obtained for the axisymmetric
potential. It turned out that including the bar changed the
resonance frequency, on average, by 0.4\%. For example, for one of
the stars from the Hercules stream with coordinates
$(x,y,z,U,V,W)=(0.041, -0.031, -0.128, -42.75, -56,70, -6.26)$
(the positions in kpc, the velocities in km s$^{-1}$), the
resonance frequency was found to be 52.74 and 52.91 km s$^{-1}$
kpc$^{-1}$ in the case of the axisymmetric potential and with the
inclusion of the bar potential, respectively. The orbits found for
this star are shown in Fig.~2. As can be seen from this figure,
the orbits in the rotating bar frame obtained in these two cases
differ visually, but not much. This is because the gravitational
contribution of the bar to the total Galactic potential is small
(Fern\'andez et al. 2008). At the same time, however, the orbit
acquires the property of stochasticity. As can be seen from Fig.
2, the orbit under the influence of the bar (right panel) becomes
more blurred (because it undergoes libration near the resonance)
than the periodic resonant orbit obtained in the axisymmetric
potential (left panel). Thus, the orbit becomes quasi-periodic
under the influence of the bar. As our integration showed, the
filling of the plane in the second case slowly but expands with
increasing integration time. For example, the number of ones $K$
in the $(X_p,Y_p)$ coordinate plane, in our case of size
$512\times512$, evolves as follows: $K=838, 1398, 1916,$ and 2378
at the integration time $T=1100, 2200, 3300,$ and 4400~Myr,
respectively.

As a result, it turned out that when the bar was included, the
resonance frequencies changed by a negligibly small value, while
the shape and orientation of orbits in the bar frame did not
change. Thus, we showed that to solve our problem of determining
the bar pattern speed and orientation, it would suffice to study
the periodic orbits obtained in the axisymmetric potential. This
is important, because the bar potential is not known with
certainty; there are a multitude of its models in the literature.
We only know that its gravitational contribution to the total
Galactic potential is small.

The influence of the bar was studied in detail using simulations
by Fux (2001). We do not set such a goal in this paper. We need to
solve the inverse problem: to estimate the parameters of the bar
by assuming that the features of the Hercules and Wolf 630 streams
are produced by its long-term dynamical effect.

The bar orientation in the Galactic $(X,Y)$ plane can be
determined from the orientation of orbits in the $(X_p,Y_p)$
plane. Indeed, while in the axisymmetric case the orientation of
the resonant orbits is arbitrary, the virtual introduction of a
bar will retain only those orbits which are reflection-symmetric
with respect to at least one of the bar principal axes (Fux 2001).

 \begin{figure} {\begin{center}
 \includegraphics[width=120mm]{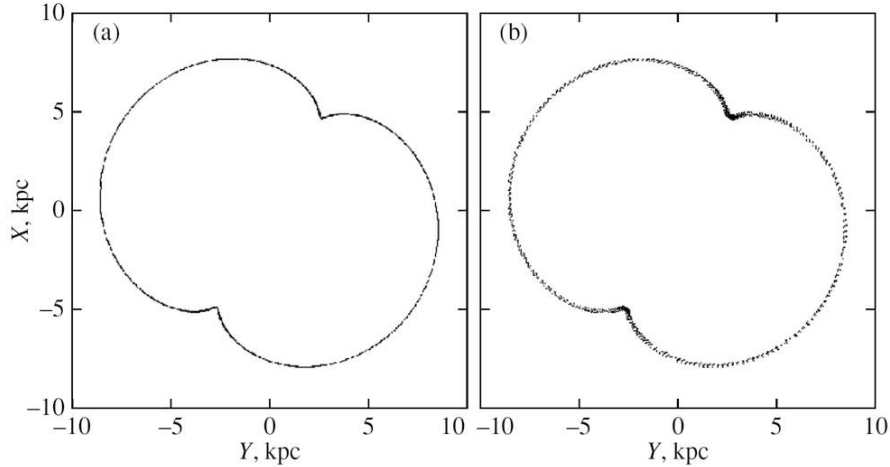}
 \caption{Example of orbit integration in a coordinate system rotating with
the bar pattern speed in the axisymmetric potential (a) and with
the addition of the bar potential (b); the Galactic center lies at
the coordinate origin; the Sun’s coordinates are
$(X,Y)=(8.5,0)$~kpc.}
 \label{f333} \end{center} } \end{figure}
 \begin{figure} {\begin{center}
 \includegraphics[width=120mm]{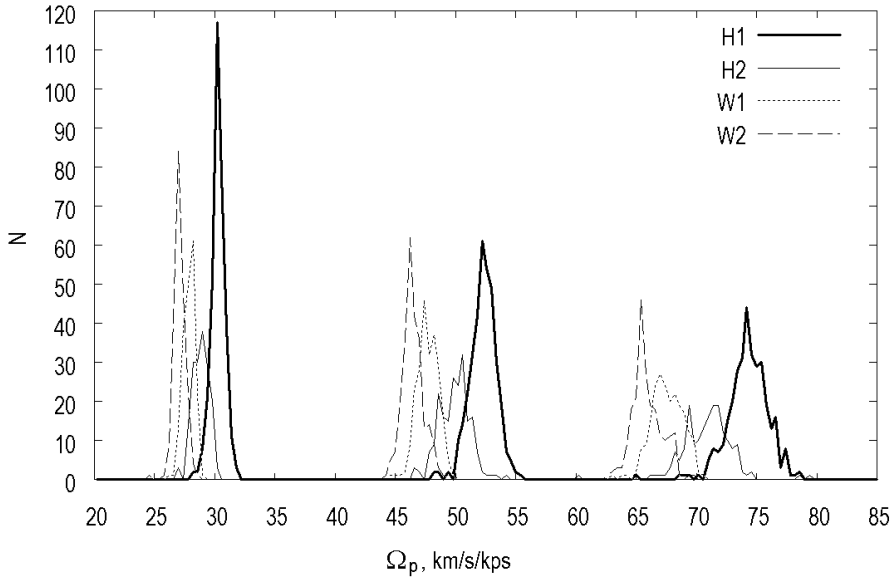}
 \caption{Distributions of the number of stars $N$ (the number of periodic orbits)
 as a function of the resonance frequency $\Omega_p$.}
 \label{f2} \end{center} } \end{figure}
 \begin{figure} {\begin{center}
 \includegraphics[width=160mm]{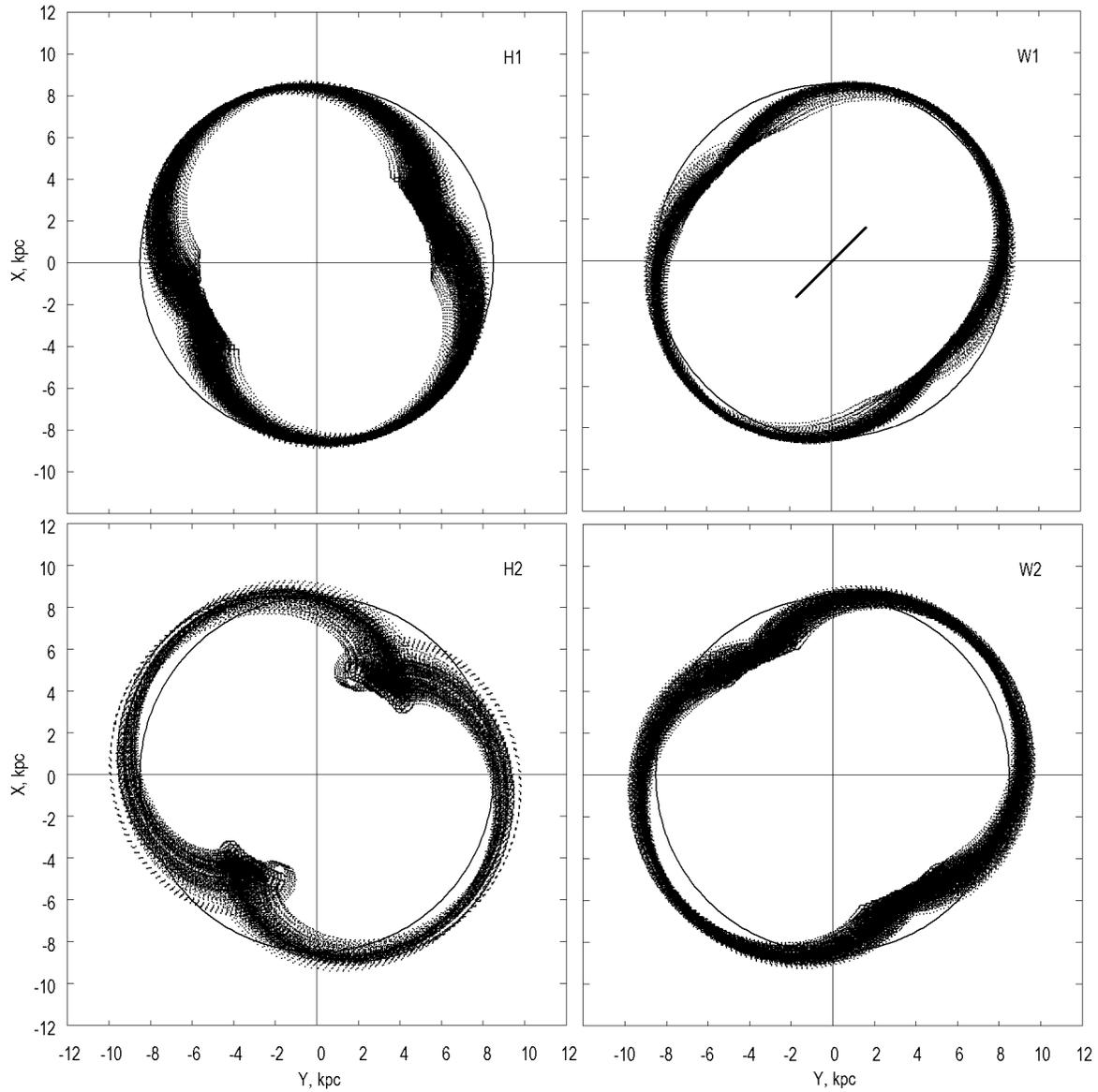}
 \caption{Examples of resonant periodic orbits in a coordinate system
rotating with the bar pattern speed. Each graph presents a
circumference of radius $R_0=8.5$~kpc; the Galactic center lies at
the coordinate origin; the Sun's coordinates are
$(X,Y)=(8.5,0)$~kpc; the upper right graph schematically shows the
bar oriented at an angle of about $45^\circ$ to the Sun's
direction.}
 \label{f3} \end{center} } \end{figure}
 \begin{figure} {\begin{center}
 \includegraphics[width=80mm]{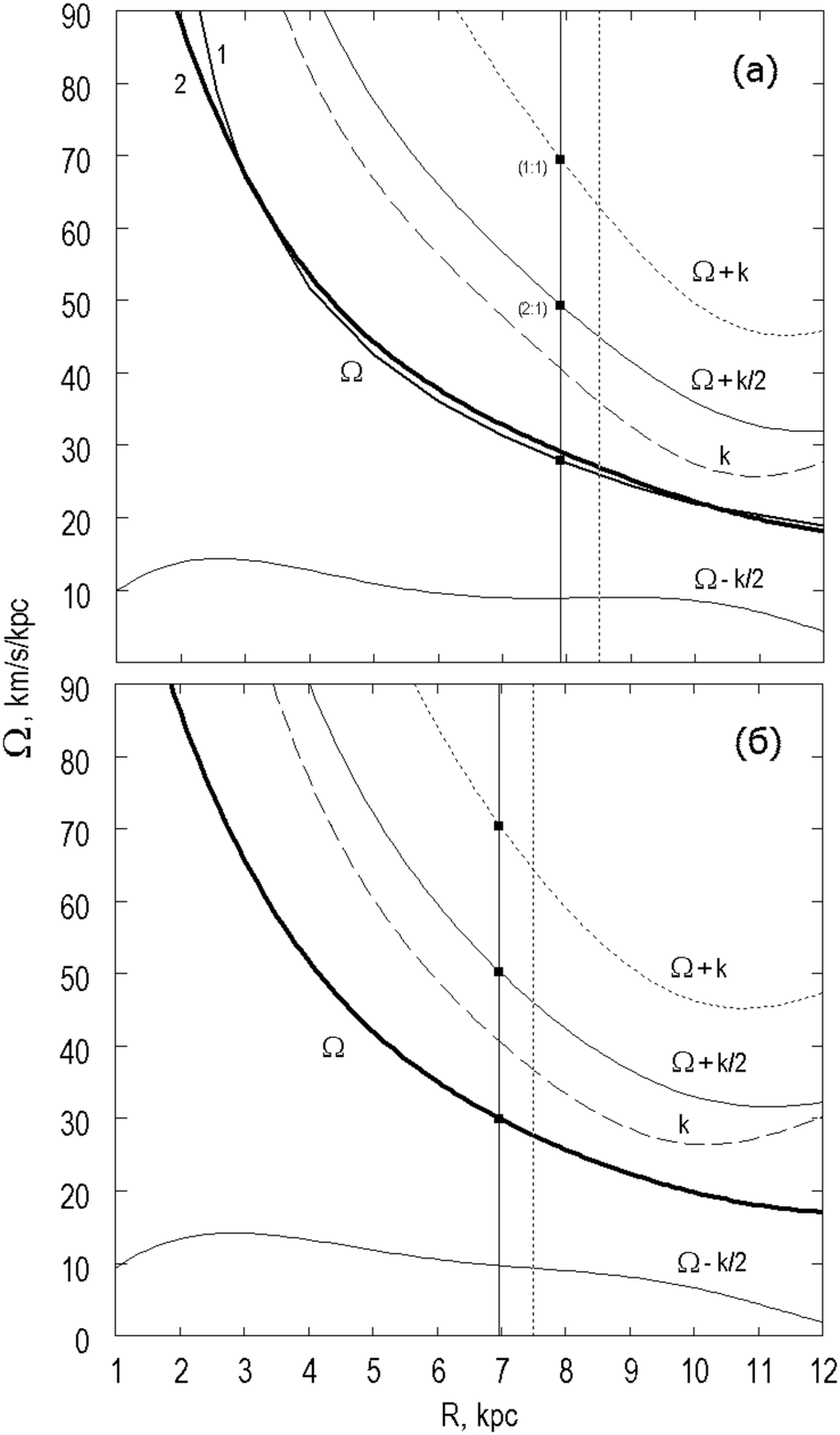}
 \caption{Galactic rotation curve constructed with
 $R_0=8.5$ (a) and $7.5$ (b) kpc; the Sun’s position on both graphs is marked by
the vertical dotted line, $R$ is the Galactocentric distance of
the star.}
 \label{f4} \end{center} } \end{figure}

\section*{RESULTS}
Using the above-described method of searching for periodic orbits
in an axisymmetric potential for each star of the identified H1,
H2, W1, and W2 features, we found the resonance frequencies
$\Omega_c$, $\Omega_{1:1}$ and $\Omega_{2:1}$. The interval of
integration was 4~Gyr.

The corresponding histograms, i.e., the distributions of the
number of stars $N$ as a function of the resonance frequency, are
presented in Fig.~3. The histograms occupying the left, right, and
central parts of the graph refer to the corotation, 1:1, and 2:1
resonance frequencies, respectively. We associate the bar pattern
speed below denoted by $\Omega_{bar}$ with the 2:1 resonance
frequency. The individual distributions are seen to be Gaussian.
The 1:1 resonant and corotation resonant orbits exhibit the
largest and smallest dispersions, respectively. The 2:1 resonant
orbits, in accordance with Eq.~(10), are intermediate between
them.

Figure 4 presents the families of orbits in a coordinate system
rotating with the bar pattern speed. The orbits of stars in the H1
and H2 streams are elongated perpendicularly to the bar major
axis, while the orbits of stars in the W1 and W2 streams are
elongated along the bar. As can be seen from Fig.~4, the bar angle
$\theta_{bar}$ lies within the range $40^\circ-60^\circ$. Thus,
the bar angle is determined by the inclination of the orbital
major axis, as shown in Fig. 4 (upper right panel). In accordance
with the central limit theorem, we assume that the deviations of
the orbital inclinations of stars from the samples considered obey
a normal law. The bar angle for each sample was calculated as the
mean of the orbital inclinations of stars from the samples; the
error in the angle was calculated as the root--mean--square
deviation from the mean.

The quantitative estimates of $\Omega_{bar}$ and $\theta_{bar}$
found from the results presented in Figs. 3 and 4 are given in
Table 2. As can be seen from the table, the bar pattern speed
$\Omega_{bar}$ and the bar angle $\theta_{bar}$ determined from
the pair of W2 and H1 have larger discrepancies between themselves
than those from the pair of W1 and H2.

Let us show that our estimates are robust when using the various
Galactic rotation curves constructed with the constant $R_0$ whose
values lie within a fairly wide range, from 7.5 to 8.5 kpc.

Figure 5a provides the Galactic rotation curves no.~1 from the
tabulated data of Allen and Santill\'an (1991) and no.~2 from the
same initial data that were used in Bobylev et al. (2008) but
constructed with $R_0=8.5$~kpc.

Bobylev et al. (2008) used data on the three dimensional space
velocity field of young open star clusters and the radial
velocities of HI clouds and HII regions. The rotation curve no. 2
was determined by using the first six terms of the Taylor
expansion of the angular velocity of Galactic rotation $\Omega$ in
Bottlinger’s equations. The terms of the series found at
$R_0=8.5$~kpc are
 $\Omega_0  =-26.96\pm0.58$~km s$^{-1}$ kpc$^{-1}$,
 $\Omega^1_0=  3.52\pm0.07$~km s$^{-1}$ kpc$^{-2}$,
 $\Omega^2_0= -0.470\pm0.046$~km s$^{-1}$ kpc$^{-3}$,
 $\Omega^3_0=  0.042\pm0.032$~km s$^{-1}$ kpc$^{-4}$,
 $\Omega^4_0= -0.218\pm0.019$~km s$^{-1}$ kpc$^{-5}$,
 $\Omega^5_0=  0.108\pm0.013$~km s$^{-1}$ kpc$^{-6}$,
 where $\Omega_0,\Omega^1_0,...,\Omega^5_0 $ are the
angular velocities and the corresponding derivatives at $R=R_0,$
and $R$ is the Galactocentric distance of the star. The Oort
constants are
 $A= 14.94\pm0.28$ km s$^{-1}$ kpc$^{-1}$ and
 $B=-12.02\pm0.65$ km s$^{-1}$ kpc$^{-1}$.

It can be seen from Fig. 5a that curves 1 and 2 have significant
differences only at $R<2.5$~kpc, which is not critical for our
goals. Curve 2 has an analytical representation, which allows the
dependences
 $k(R)=2\Omega(1+1/2\cdot R/\Omega\cdot d\Omega/dR)^{1/2}$,
 $\Omega+k/2$,
 $\Omega-k/2,$ and
 $\Omega+k$
shown in Fig. 5 to be calculated. The vertical dotted line in Fig.
5a marks the distance $R_0=8.5$~kpc, the thin vertical line
indicates the radius of the outer Lindblad resonance (OLR)
$R_{OLR}=7.9$~kpc (we found $R_{OLR}=0.93 R_0$ based on the data
in Fig. 3 from the mean of the four corresponding frequencies),
the three resonances of interest to us are marked: the corotation,
(2:1), and (1:1) ones. According to present-day estimates, the
exact value of R0 lies within the range 7.1--8.5~kpc (see, e.g.,
Table 1 from Foster and Cooper (2010)). Therefore, it is of
interest to trace whether the positions of the resonances change
with $R_0.$ Figure 5b provides the rotation curve constructed
directly from the data of Bobylev et al. (2008) at $R_0=7.5$~kpc.
As can be seen from Fig.~5, when $R_0$ decreases by 1~kpc, the
resonance frequencies remain almost unchanged at
$R_{OLR}=0.93R_0.$ As we see from the graphs, the bar pattern
speed corresponding to the 2:1 resonance frequency is about 52 km
s$^{-1}$ kpc, consistent with the results that we have obtained
above by an independent method.

 {\begin{table}[t]                            
 \caption[]
 {\small\baselineskip=1.0ex
 Parameters found from the $H_{1,2}$ and $W_{1,2}$ features  }
 \label{t:2}
 \begin{center}\begin{tabular}{|c|c|c|}\hline
 Feature & $\Omega_{bar},$ km s$^{-1}$ kpc$^{-1}$ & $\theta_{bar},$ deg\\\hline
 $H_1$ & $52.3\pm1.1$ & $62\pm11$ \\
 $H_2$ & $49.9\pm1.2$ & $51\pm9~$ \\
 $W_1$ & $47.7\pm0.9$ & $45\pm10$ \\
 $W_2$ & $46.4\pm0.9$ & $38\pm9~$ \\\hline
 Mean & $49.1\pm1.3$ & $49\pm5~$ \\\hline
 \end{tabular}\end{center}\end{table}}
{\begin{table}[t]                                                
 \caption[]{\small\baselineskip=1.0ex\protect
 Parameters of the Galactic ``short'' (the upper part of the table) and
 ``long'' (the lower part of the table) bars }
 \begin{center}
 \label{t:liter}
 \begin{tabular}{|c|c|c|r|c|r|r|r|r|r|r|}\hline
       $\Omega_{bar},$         & $R_{bar},$  & $\theta_{bar}$  &  Source \\
    km s$^{-1}$ kpc$^{-1}$     &   kpc       &                 &         \\\hline
        63       & $2.4\pm0.5$ &   $16^\circ\pm2^\circ$ &      Binney et al. (1991) \\
  $53\pm3$       &     ---     &                  ---   &             Dehnen (1999) \\
  $59\pm5$       &     ---     &                  ---   &  Debattista et al. (2002) \\
  $60\pm5$       & $3.4\pm0.3$ &    $20^\circ-25^\circ$ &    Bissantz et al. (2003) \\
  ---            &     ---   & $11.1^\circ\pm0.7^\circ$ &       Robin et al. (2003) \\
  ---            &   $2.5-3.0$ & $22^\circ\pm5.5^\circ$ & Babusiaux and Gilmore (2005) \\
  ---            &     ---     &    $20^\circ-35^\circ$ & L\'opez-Corredoira et al. (2005) \\
  ---            &   ---     & $12.6^\circ\pm3.2^\circ$ & Cabrera-Lavers et al. (2007) \\
 $57.4^{+2.8}_{-3.3}$ &   ---  &     $0^\circ-30^\circ$ &       Chakrabarty (2007) \\
  ---            &     ---     &    $24^\circ-27^\circ$ & Rattenbury et al. (2007) \\
  $57.7$         &     ---     &                  ---   &   Gardner and Flinn (2010) \\
  ---            &    $3$      &    $20^\circ-25^\circ$ &    Bobylev et al. (2014) \\
  $56\pm2$       &     ---     &                  ---   &     Antoja et al. (2014) \\\hline

  ---            &   ---       &             $44^\circ$ &        Sevenster et al. (1999) \\
  ---            & $4.4\pm0.5$ &  $44^\circ\pm10^\circ$ &        Benjamin et al. (2005) \\
  ---            &   ---       &  $43^\circ\pm17^\circ$ & Groenewegen and Blommaert (2005) \\
  ---            & $3.9$       &  $43^\circ\pm 7^\circ$ & L\'opez-Corredoira et al. (2007) \\
  ---            & $4$       & $43.0^\circ\pm1.8^\circ$ &   Cabrera-Lavers et al. (2007) \\
  $56.7$         &     ---     &                  ---   &         Gardner and Flinn (2010) \\
  $49.1\pm1.3$   & $4.2\pm0.7$ & $49^\circ\pm5^\circ$   &         this paper \\
 \hline
 \end{tabular}\end{center}
 \end{table}}

\section*{DISCUSSION}
Let us first discuss some of the known bar characteristics.
Table~3 gives the estimates of the bar pattern speed, size, and
orientation obtained by different authors using different data.
The characteristics of both classical ``short'' and ``long'' bars
are given.

Binney et al. (1991) and Bissantz et al. (2003) estimated the
parameters by comparing the radio observations of central Galactic
regions (HI, CO) with various bar models.

Debattista et al. (2002) estimated the bar pattern speed from
radio observations of 250 OH/IR stars. These are oxygen-rich stars
with envelopes at the asymptotic-giant-branch phase with strong
mass loss. The observations yielded  $\Delta V\equiv\Omega_{bar}
R_0-V_{LSR}=252\pm41$~km s$^{-1}$, while $\Omega_{bar}$ was
calculated at $R_0=8$~kpc and $V_{LSR}=220$ km s$^{-1}$. Thus,
this is a rare example of a direct measurement of the bar pattern
speed. The same stars were also used in Sevenster et al. (1999).

Benjamin et al. (2005) estimated the size and orientation of the
long bar by analyzing the star counts based on the GLIMPSE
(Galactic Legacy Infrared Mid-Plane Survey Extraordinaire)
infrared survey produced on the basis of observations with the
Spitzer Space Telescope. For these purposes, Groenewegen and
Blommaert (2007) used Mira variables from the OGLE-II survey
(Wozniak et al. 2002), designed to search for microlensing
effects. The arguments for the hypothesis of a long bar were also
presented by L\'opez-Corredoira et al. (2007). Shortly afterwards,
based on the counts of stars from the 2MASS catalog, this group of
authors (Cabrera-Lavers et al. 2007) proposed a two-bar model.

A similar method of star counts was applied using red-giant-clump
stars from the CIRSI (Cambridge InfraRed Survey Instrument) survey
by Babusiaux and Gilmore (2005), from the 2MASS catalog by
L\'opez-Corredoira et al. (2005), and from the OGLE-II survey by
Rattenbury et al. (2007). Robin et al. (2003) extracted
information about the bar size and orientation by analyzing the
data from the Hipparcos catalogue (1997).

Dehnen (1999), Chakrabarty (2007), and Gardner and Flinn (2010)
estimated the bar parameters based on simulations to explain the
observed bimodal $UV$ velocity distribution of solar-neighborhood
stars. It is important to note that one of the models in Gardner
and Flinn (2010) contained two bars for which close pattern speeds
were found. This means that the two bars can rotate almost
synchronously.

Bobylev et al. (2014) estimated the bar orientation parameters
using the XPM catalog (Fedorov et al. 2010), which contains the
proper motions of millions of stars. There are photometric data
from the 2MASS catalog for these stars, which allowed the
photometric distances to be estimated, while using the proper
motions allowed one to eliminate the foreground stars and
literally to ``see'' the bar.

Antoja et al. (2014) found the bar pattern speed by comparing
their analytical bar model with stars from the RAVE4 catalog.
Distance estimates and highly accurate radial velocities are
available for these stars. In the model of these authors, the
Hercules stream was assumed to be produced by the influence of the
bar and the Galactic spiral structure.

Table 3 does not encompass all of the known results. Note that
quite a large selection of Galactic bar parameter determinations
can be found in Vanhollebeke et al. (2009).

The resonant orbits we found show that the bar pattern speed
$\Omega_{bar}$ lies within the range 45--55 km s$^{-1}$
kpc$^{-1}$, while the bar angle $\Omega_{bar}$ is within the range
$40^\circ-60^\circ.$ The results obtained are consistent with the
view that the Hercules and Wolf~630 streams could be formed by a
single mechanism associated with the splitting of the $UV$
velocity plane under the influence of the Galactic bar.

Our results are in agreement with the simulations of Dehnen (1999)
for $R_{OLR}\approx0.9R_0.$ The bar size estimated from the
rotation curve by taking into account the confidence region for
$\Omega_{bar}$ is $R_{bar}=4.2\pm0.7$~kpc (for the adopted
$R_0=8$~kpc). As can be seen from Table~3, this resembles a long
bar.

\section*{CONCLUSIONS}
We analyzed the four most significant features in the $UV$
velocity distribution of solar-neighborhood stars: H1, H2 from the
Hercules stream and W1, W2 from the Wolf 630 stream with the
$(U,V)$ coordinates of their centers $(-33,-51), (-71,-48),
(21,-26),$ and $(40,-24)$ km s$^{-1}$, respectively. Based on the
assumption that the Hercules and Wolf 630 streams were induced by
the central Galactic bar, we formulated the problem of determining
the bar characteristics independently from each of the identified
features.

To construct the Galactic orbits of the individual stars forming
these streams, we used the model of Allen and Santill\'an (1991).
From the set of constructed orbits, we selected only the stable
orbits in resonance with the bar. Analysis of the resonant orbits
found showed that the bar pattern speed $\Omega_{bar}$ lies within
the range 45--55 km s$^{-1}$ kpc$^{-1}$ with a mean of
$49.1\pm1.3$ km s$^{-1}$ kpc$^{-1}$, while the bar angle is within
the range $40^\circ-60^\circ$ with a mean of $49\pm5^\circ.$ These
estimates were shown to be robust when using the various Galactic
rotation curves constructed with the constant $R_0$ whose values
lie within a fairly wide range, from 7.5 to 8.5 kpc.

The results obtained are consistent with the view that the
Hercules and Wolf 630 streams could be formed by a single
mechanism associated with the fact that a long-term influence of
the Galactic bar led to a characteristic bimodal splitting of the
$UV$ velocity plane.

\subsection*{ACKNOWLEDGMENTS}
We are grateful to the referees for their helpful
remarks that contributed to an improvement of this paper. This
work was supported by the ``Transitional and Explosive Processes
in Astrophysics'' Program P--41 of the Presidium of Russian
Academy of Sciences.

 \bigskip{REFERENCES}\medskip
 {\small

1. C. Allen and A. Santill\'an, Rev. Mex. Astron. Astrof. 22, 255
(1991).

2. T. Antoja, F. Figueras, D. Fern\'andez, and J. Torra, Astron.
Astrophys. 490, 135 (2008).

3. T. Antoja, O. Valenzuela, B. Pichardo, E. Moreno, F. Figueras,
and D. Fern\'andez, Astrophys. J. 700, L78 (2009).

4. T. Antoja, A. Helmi, W. Dehnen, O. Bienaym\'e, J.
Bland-Hawthorn, B. Famaey, K. Freeman, B.K. Gibson, et al.,
Astron. Astrophys. 563, 60 (2014).

5. R. Asiain, F. Figueras, J. Torra, and B. Chen, Astron.
Astrophys. 341, 427 (1999).

6. C. Babusiaux and G. Gilmore, Mon. Not. R. Astron. Soc. 358,
1309 (2005).

7. R.A. Benjamin, E. Churchwell, B.L. Babler, R. Indebetouw, M.R.
Meade, B.A. Whitney, C. Watson, M.G. Wolfire, et al., Astrophys.
J. 630, L149 (2005).

8. T. Bensby, M.S. Oey, S. Feltzing, and B. Gustafsson, Astrophys.
J. 655, L89 (2007).

9. J. Binney, O.E. Gerhard, A. Stark, J. Bally, and K.I. Uchida,
Mon. Not. R. Astron. Soc. 252, 210 (1991).

10. N. Bissantz, P. Englmaier, and O. Gerhard, Mon. Not. R.
Astron. Soc. 340, 949 (2003).

11. V.V. Bobylev, G.A. Gontcharov, and A.T. Bajkova, Astron. Rep.
50, 733 (2006).

12. V.V. Bobylev and A.T. Bajkova, Astron. Rep. 51, 372 (2007).

13. V.V. Bobylev, A.T. Bajkova, and A.S. Stepanishchev, Astron.
Lett. 34, 515 (2008).

14. V.V. Bobylev, A.T. Bajkova, and A.A. Myll\"ari, Astron. Lett.
36, 27 (2010).

15. V.V. Bobylev, A.V. Mosenkov, A.T. Bajkova, and G.A.
Gontcharov, Astron. Lett. 40, 86 (2014).

16. J. Bovy, Astrophys. J. 725, 1676 (2010).

17. E.J. Bubar and J.R. King, Astron. J. 140, 293 (2010).

18. A. Cabrera-Lavers, P.L. Hammersley, C. Gonz\'alez-Fern\'andez,
M. L\'opez-Corredoira, F. Garz\'on, and T.J. Mahoney, Astron.
Astrophys. 465, 825 (2007).

19. D. Chakrabarty, Astron. Astrophys. 467, 145 (2007).

20. E. Chereul, M. Cr\'ez\'e, and O. Bienaym\'e, Astron.
Astrophys. 340, 384 (1998).

21. V.P. Debattista, O. Gerhard, and M.N. Sevenster, Mon. Not. R.
Astron. Soc. 334, 355 (2002).

22. W. Dehnen, Astron. J. 115, 2384 (1998).

23. W. Dehnen, Astrophys. J. 524, L35 (1999).

24. W. Dehnen, Astron. J. 119, 800 (2000).

25. O.J. Eggen, Publ. Astron. Soc. Pacif. 81, 553 (1969).

26. B. Famaey, A. Jorissen, X. Luri, M. Mayor, S. Udry, H.
Dejonghe, and C. Turon, Astron. Astrophys. 430, 165 (2005).

27. P.N. Fedorov, V.S. Akhmetov, V.V. Bobylev, and A.T. Bajkova,
Mon. Not. R. Astron. Soc. 406, 1734 (2010).

28. D. Fern\'andez, F. Figueras, and J. Torra, Astron. Astrophys.
480, 735 (2008).

29. T. Foster and B. Cooper, ASP Conf. Ser. 438, 16 (2010).

30. C. Francis and E. Anderson, New Astron. 14, 615 (2009).

31. R. Fux, Astron. Astrophys. 373, 511 (2001).

32. E. Gardner and C. Flinn, Mon. Not. R. Astron. Soc. 405, 545
(2010).

33. G.A. Gontcharov, Astron. Lett. 32, 795 (2006).

34. M.A.T. Groenewegen and J.A.D.L. Blommaert, Astron. Astrophys.
443, 143 (2005).

35. The Hipparcos and Tycho Catalogues, ESA SP-1200 (1997).

36. F. van Leeuwen, Astron. Astrophys. 474, 653 (2007).

37. M. L\'opez-Corredoira, A. Cabrera-Lavers, and O.E. Gerhard,
Astron. Astrophys. 439, 107 (2005).

38. M. L\'opez-Corredoira, A. Cabrera-Lavers, T.J. Mahoney, P.L.
Hammersley, F. Garz\'on, and C. Gonz\'alez-Fern\'andez, Astron. J.
133, 154 (2007).

39. V.V. Orlov and N.Ya. Sotnikova, in Astronomy: Traditions,
Present, Future, Ed. by V.V. Orlov, I.P. Reshetnikov, and N.Ya.
Sotnikova (SPb Gos. Univ., St. Petersburg, 2007), p. 169 [in
Russian].

 40. J. Palou$\breve{s}$, B. Jungwiert, and J. Kopeck$\acute{y}$, Astron.
Astrophys. 274, 189 (1993).

41. L. Pompeia, T. Masseron, B. Famaey, S. van Eck, A. Jorissen,
I. Minchev, A. Siebert, C. Sneden, et al., Mon. Not. R. Astron.
Soc. 415, 1138 (2011).

42. A.C. Quillen and I. Minchev, Astron. J. 130, 576 (2005).

43. N.J. Rattenbury, S. Mao, T. Sumi, and M.C. Smith, Mon. Not. R.
Astron. Soc. 378, 1064 (2007).

44. A.C. Robin, C. Reyl\'e, S. Derri\'ere, and S. Picaud, Astron.
Astrophys. 409, 523 (2003).

45. R. Sch\"onrich, J. Binney, and W. Dehnen, Mon. Not. R. Astron.
Soc. 403, 1829 (2010).

46. M. Sevenster, P. Saha, D. Valls-Gabaud, and R. Fux, Mon. Not.
R. Astron. Soc. 307, 584 (1999).

47. R.S. de Simone, X. Wu, and S. Tremain, Mon. Not. R. Astron.
Soc. 350, 627 (2004).

48. J. Skuljan, J.B. Hearnshaw, and P.L. Cottrell, Mon. Not. R.
Astron. Soc. 308, 731 (1999).

49. E. Vanhollebeke, M.A.T. Groenewegen, and L. Girardi, Astron.
Astrophys. 498, 95 (2009).

50. P.R. Wozniak, A. Udalski, M. Szymanski, M. Kubiak, G.
Pietrzynski, I. Soszynski, and K. Zebrun, Acta Astron. 52, 129
(2002).

 }
\end{document}